\documentstyle{article}

\newcommand{\bea}{\begin{eqnarray}}
\newcommand{\eea}{\end{eqnarray}}
\newcommand{\beq}{\begin{equation}}
\newcommand{\eeq}{\end{equation}}
\newcommand{\nn}{\nonumber}

\newfont{\cms}{cmss8 scaled 1440}

\def\/{\over}
\begin{document}

\parindent=1 em
\frenchspacing

\noindent
\begin{minipage}[t]{7cm}\rule{70mm}{1mm} \\[2mm]
{\cms  Konstanz University \hfill $\bullet$  \hfill Theory Group
  \hfill \\[4mm]
$\bigl\langle$ \hfill Gravity \hfill $\bigr|$ \hfill Quantum
   Theory \hfill $\bigr|$ \hfill
   Optics \hfill $\bigr\rangle$} \hfill \\[1mm]
\rule{70mm}{1mm}
\end{minipage}
\hfill \\[1.25cm]
\noindent{\em Preprint  KONS-RGKU-95-01 --  quant-ph/9503012}\\[2cm]

\begin{center} {\LARGE \bf
Relation between energy shifts and relaxation rates for a small system coupled
to a reservoir}
\\[1cm]
{\bf
J\"urgen Audretsch\footnote{e-mail: Juergen.Audretsch@.uni-konstanz.de},
 Rainer M\"uller\footnote{e-mail:
Rainer.Mueller@.uni-konstanz.de \hfill{\normalsize\it to appear in Physics
Letters A}\quad}
 and Markus Holzmann
  \\[0.3cm]
\normalsize \it Fakult\"at f\"ur Physik der Universit\"at Konstanz\\
\normalsize \it Postfach 5560 M 674, D-78434 Konstanz, Germany}
\vspace{0.7cm}

\begin{minipage}{15cm}
\begin{abstract}
For a small system the coupling to a reservoir causes energy shifts
as well as transitions between the system's energy levels. We
show for a general stationary situation that the energy shifts can
essentially be reduced to the relaxation rates.
The effects of reservoir fluctuations and self reaction are treated
separately. We apply the results to a two-level atom coupled to a
reservoir which may be the vacuum of a radiation field.
\end{abstract}
\end{minipage}
\end{center}

\vspace{0.5cm}


\section{Introduction}

For a small quantum system which is coupled to a reservoir,
energy shifts due to the coupling will occur as well as transitions
between the system's energy levels. Our aim in the present letter is to derive
relations between these relaxation rates and energy shifts
which are valid under very general conditions. We consider a system which
moves in a stationary way on a possibly accelerated trajectory
in a general spacetime (which may be curved or possess nontrivial
boundaries). The stationarity of the situation demands that the system moves
along the orbits of a timelike Killing vector field.
Stationarity is also assumed for the reservoir. The case of
an extended system at rest is included.
Typical realizations of such a situation are an atom
or an elementary particle (system) coupled to a quantized radiation field
in the vacuum or a many-photon state (reservoir), whereby the atom may
be accelerated. In this case the energy shift is the Lamb shift.

For the description of the system-reservoir interaction we will
generalize the formalism which has been established by Dalibard,
Dupont-Roc and
Cohen-Tannoudji (DDC) \cite{Dalibard84} and extended in
\cite{Audretsch94a,Audretsch94b} to the situation just described.
It leads to a clear and physically appealing interpretation of
the processes in terms of the underlying physical mechanisms
since it allows the separate discussion of the effects of
reservoir fluctuations and self reaction (or radiation reaction).

\section{Reservoir fluctuations and self reaction}

First we provide the necessary general relations.
We consider a spacetime which is covered by a coordinate system
$x=(t,\vec x)$. The time coordinate $t$ is assumed to be the
``natural'' time for the description of the reservoir. The system
is described by a stationary (accelerated) trajectory $x(\tau)=
(t(\tau),\vec x(\tau))$. This may be realized for example by a
pointlike object moving on this trajectory. The respective proper time
variable $\tau$ is used for the parametrization of the trajectory.

The time evolution of system and reservoir has to be specified
in terms of a single time variable for which we choose
the system time $\tau$.
The Hamiltonian which governs the dynamics with respect to
$\tau$ is given by
\beq H=H_S(\tau)+H_R(t) {dt\/ d \tau} + V(x(\tau))  \label{eq1}\eeq
where $H_S$ is the free Hamiltonian of the system. $H_R(t)$ is the
free reservoir Hamiltonian with respect to the time variable $t$.
The factor $dt/d \tau$ in (\ref{eq1}) is due to the corresponding
change of variables. $V(x(\tau))$ represents the coupling between
system and reservoir. It is effective only on the trajectory
$x(\tau)$ of the system. It is assumed to have the general linear form
\beq V(x(\tau))= - g \sum_i R_i(\tau)S_i(x(\tau)) \label{eq2}\eeq
with $g$ being a coupling constant. $R_i$ and $S_i$ are hermitean
reservoir and system operators. For an extended system at rest, we can
set $\tau=t$ and replace $x(\tau)$ by $t$ here and in the equations
below.

We can now write down the Heisenberg equations of motion for
observables of the system and the reservoir. For our purposes,
the following will be important: It is possible in the solution of the
Heisenberg equation for the reservoir variable $R_i$ to distinguish
on one hand the part which is present even in the absence
of the coupling. It is independent of the system and is called
the {\it free part} $R_i^f$ of $R_i$. The remaining contribution
is caused by the presence of the system and contains the coupling
constant $g$. It is called the {\it source part} $R_i^s$ of $R_i$:
\beq R_i(\tau)=R_i^f(\tau) +R_i^s(\tau). \label{eq3}\eeq

We consider the rate of change of an arbitrary system variable $G$.
Because of the coupling (\ref{eq2}), reservoir operators appear in
the Heisenberg equations of $G$. According to (\ref{eq3}), they
can be divided into their free and source parts. The rate of change
of $G$ due to the coupling can therefore be split into two
contributions which correspond to two different physical mechanisms:
(i) the change in $G$ produced by the fluctuations of the reservoir
which are present even in the absence of the system -- this portion is related
to the free part of the reservoir and is called the contribution
of {\it reservoir fluctuations} to $dG/ d \tau$ -- and (ii) the
change in $G$ due to the interaction with the excitations of the
reservoir which are caused by the system itself. This is the
contribution of {\it self reaction} or radiation reaction and is
connected with the source part of the field.
Following DDC \cite{Dalibard82}, we adopt a symmetric ordering between
system and reservoir operators.

In a perturbative approach, we take into account only terms up
to second order in $g$. Since we are interested only in the
dynamics of the system, we average over the reservoir degrees
of freedom. We assume that the density matrix is factorized
into a system and a reservoir part at the initial time $\tau=0$:
$\rho (0)=\rho_S(0)\, \rho_R(0)$. We
select one specific system state $|a \rangle$ and take the expectation
value with respect to that state.

Proceeding essentially as in \cite{Dalibard84}, we find
for the contribution of reservoir fluctuations
\beq \left\langle {dG \over {d\tau}} (\tau)\right\rangle_{rf} =
  i \left\langle \left[ H^{eff}_{rf}(\tau) , G(\tau) \right]
  \right\rangle - {g^2 \over 2} \sum_i \left\langle \left[ Y_i(\tau),
  [S_i^f(\tau),G(\tau)] \right] + \left[ S_i^f(\tau), [Y_i(\tau),G(\tau)]
  \right] \right\rangle \label{eq9}\eeq
whereas the contribution of self reaction is (curly brackets denote the
anticommutator):
\beq \left\langle {dG \over {d\tau}} (\tau)\right\rangle =
  i \left\langle \left[ H^{eff}_{sr}(\tau), G(\tau) \right] \right\rangle
  - {g^2 \over 2} \sum_i \left\langle \left\{ Z_i(\tau), [S_i^f(\tau),G(\tau)]
  \right\} - \left\{ S_i^f(\tau), [Z_i(\tau),G(\tau)] \right\}
  \right\rangle. \label{eq10}\eeq
In (\ref{eq9}) and (\ref{eq10}), brackets $\langle \dots \rangle$
denote averaging over
the reservoir and taking the expectation value in the system state
$|a \rangle$. Furthermore we have introduced the effective Hamiltonians
\beq  H^{eff}_{rf}(\tau) :={ig^2 \over 2} \sum_i \left[ Y_i(\tau),S_i^f(\tau)
\right], \qquad H^{eff}_{sr}(\tau) :=-{ig^2 \over 2} \sum_i \left\{
Z_i(\tau),S_i^f(\tau) \right\} \label{eq1112}\eeq
with
\beq Y_i(\tau) :=\sum_j \int_{0}^{\infty}d\tau' \,
  C_{ij}^R (x(\tau),x(\tau')) S_j^f(\tau'), \qquad
Z_i(\tau) :=\sum_j \int_{0}^{\infty}d\tau' \,\chi_{ij}^R(x(\tau),x(\tau'))
S_j^f(\tau') \label{eq1314}\eeq
Since we are interested in time scales which are large compared with
the correlation time of the reservoir, we have extended the range
of integration in (\ref{eq1314}) to infinity.
We have also introduced the symmetric correlation function $C^R_{ij}$
and the linear susceptibility $\chi^R_{ij}$ of the reservoir:
\beq C_{ij}^R(x(\tau),x(\tau')):={1 \over 2}\, \hbox{Tr}_R \left( \rho_R
  (0)
  \left\{ R_i^f(x(\tau)),R_j^f(x(\tau')) \right\} \right) \label{eq15}\eeq
\beq \chi_{ij}^R(x(\tau),x(\tau')):={1 \over 2}\, \hbox{Tr}_R \left( \rho_R (0)
  \left[ R_i^f(x(\tau)),R_j^f(x(\tau')) \right] \right) \label{eq16}\eeq
Because of the stationarity, $C^R_{ij}$ and $\chi^R_{ij}$ are only
functions of the time difference $\tau-\tau'$.

\section{Energy shifts and relaxation rates}

The relaxation rates of the system's energy in the state $|a \rangle$ can
be obtained from (\ref{eq9}) and (\ref{eq10}) with the choice $G=H_S$.
We replace $[S_i^f,H_S^f]$ in second order by $i {d\/ d \tau} S^f_i$ and
find for the contributions of vacuum fluctuations and self
reaction to the system's relaxation rate
\beq \left\langle {dH_S \over {d\tau}}\right\rangle_{rf} =
  2ig^2\sum_{i,j}\int_{0}^{\infty}d\tau' \, C_{ij}^R(x(\tau),x(\tau'))
  {d \over {d\tau}}\chi_{ij}^S(\tau,\tau') \label{eq17}\eeq
\beq \left\langle {dH_S \over {d\tau}}\right\rangle_{sr} =
  2ig^2\sum_{i,j}\int_{0}^{\infty}d\tau' \, \chi_{ij}^R(x(\tau),x(\tau'))
  {d \over {d\tau}}C_{ij}^S(\tau,\tau'), \label{eq18}\eeq
with the symmetric correlation function and the linear susceptibility
of the system
\beq C_{ij}^S(\tau,\tau'):= {1 \over 2} \, \langle a| \left\{
  S_i^f(\tau),S_j^f(\tau') \right\} |a\rangle
=\sum_b \hbox{Re}\left( \langle a|S_i^f(0)|b \rangle \langle
  b|S_j^f(0)|a \rangle e^{i\omega_{ab}(\tau - \tau')}
  \right), \label{eq21}\eeq
\beq \chi_{ij}^S(\tau,\tau')= {1 \over 2} \, \langle a| \left[
  S_i^f(\tau),S_j^f(\tau') \right] |a\rangle
=i \sum_b \hbox{Im}\left( \langle a|S_i^f(0)|b \rangle \langle
  b|S_j^f(0)|a \rangle e^{i\omega_{ab}(\tau - \tau')}
  \right). \label{eq19}\eeq
To evaluate (\ref{eq21}) and (\ref{eq19}) we have used the stationarity of
the situation which allowed us to introduce a complete set of stationary
system states $|b \rangle$ (eigenstates of $H_S$) with energies
$\omega_{b}$ and $\omega_{ab}=\omega_a-\omega_b$.
The formulas (\ref{eq17}) and (\ref{eq18}) are the generalizations of
the corresponding equations of DDC \cite{Dalibard84} to the situation
considered here. (Note the slight differences in the definition of the
correlation functions).

Beneath that, the system-reservoir coupling leads to a shift of
the system's energy levels. The second order radiative shift of state
$|a \rangle$ is given by the expectation value of the effective
Hamiltonians (\ref{eq1112}) in that level. Again, the
total shift can be split into the contributions of vacuum
fluctuations and radiation reaction:
\beq \left(\delta E_a \right) _{rf} = -ig^2 \sum_{i,j}\int_{0}^{\infty}
  d\tau'\, C_{ij}^R(x(\tau),x(\tau'))\chi_{ij}^S(\tau,\tau') \label{eq23}\eeq
\beq\left( \delta E_a\right)_{sr} = -ig^2 \sum_{i,j}\int_{0}^{\infty}
d\tau' \, \chi_{ij}^R(x(\tau),x(\tau'))C_{ij}^S(\tau,\tau') \label{eq24}\eeq
These formulas too  are generalizations of those in \cite{Dalibard84}.
Because $C^R_{ij}$ ($\chi^R_{ij}$) is symmetric (antisymmetric) in $i$ and
$j$, terms with $\omega_{ab}=0$ do not contribute to the relaxation rates
and energy shifts. The proof, which we omit, is based on the stationarity.
Without restriction, we can therefore exclude these terms below from all
$b$ summations.

\section{Relation between energy shifts and relaxation rates}

We now come to the main point of this letter. We will prove
to second order in $g$ for a system in the state $|a \rangle$ quite
general relations between the corresponding relaxation rates (\ref{eq17})
and (\ref{eq18}) and the energy shifts (\ref{eq23}) and (\ref{eq24}).
We start with the contribution of reservoir fluctuations. Using
the explicit formula (\ref{eq19}) for the system's linear susceptibility,
the energy shift (\ref{eq23}) can be written
\beq (\delta E_a)_{rf} = -g^2  \sum_{i,j,b}\int_{0}^{\infty}
  d\tau' \, C_{ij}^R(x(\tau),x(\tau')) { d \over {d\tau}} Re \left( \langle
  a|S_i^f(0)|b \rangle \langle b|S_j^f(0)|a \rangle {1 \over
  {\omega_{ab}}} e^{i\omega_{ab}(\tau - \tau')} \right) \label{eq27}\eeq
Now we can apply the Kramers-Kronig relation \cite{Arfken85}
\beq \hbox{Re}( f(\omega_{ab})) = {1 \over \pi} \int_{-\infty}^{+\infty}
  d\omega' \, \hbox{Im}(f(\omega')) {{\cal P} \over {\omega' - \omega_{ab}}}
  \label{eq28}\eeq
to the function $f(\omega_{ab}) = \langle
  a|S_i^f(0)|b \rangle \langle b|S_j^f(0)|a \rangle \omega_{ab}^{-1}
e^{i\omega_{ab}(\tau -
\tau')}$ which is analytic for $\omega_{ab}\neq 0$ and vanishes for
$\omega_{ab} \to i \infty$. We find
\bea \left( \delta E_a \right) _{rf} &=&
   -{g^2 \over { \pi}}\sum_{i,j,b} \int_{-\infty}^{+\infty} d\omega' \,
  \int_{0}^{\infty}d\tau' \, C_{ij}^R(x(\tau),x(\tau'))
  \label{eq28a}\\
  && \qquad\times  { d \over {d\tau}} \hbox{Im} \,\left( \langle
  a|S_i^f(0)|b \rangle \langle b|S_j^f(0)|a \rangle {1 \over {\omega'}}
e^{i\omega' (\tau - \tau')} \right) {{\cal P}  \over {\omega' - \omega_{ab}}}
\nn\eea
Introducing the quantity $\Gamma_{ab}^{rf}(\omega')$ by the definition
\beq  \Gamma_{ab}^{rf}(\omega') :=
  -2 g^2\sum_{i,j}\int_{0}^{\infty}d\tau' \, C_{ij}^R(x(\tau),x(\tau'))
  {d \over {d\tau}} \hbox{Im}\left( \langle a|S_i^f(0)|b \rangle \langle
  b|S_j^f(0)|a \rangle e^{i \omega'(\tau - \tau')}\right)
  \label{eq31b}\eeq
we obtain the desired relation
\beq \left( \delta E_a \right) _{rf}= {1 \over {2 \pi}}\sum_b
  \int_{-\infty}^{+\infty} d\omega' \, {\Gamma_{ab}^{rf}(\omega') \over
  {\omega'}}  {{\cal P} \over
  {\omega' - \omega_{ab}}}.  \label{eq31}\eeq
Eq. (\ref{eq31}) connects energy shift and relaxation rate since the latter
can be expressed easily in terms of $\Gamma_{ab}^{rf}$:
\beq \left\langle {dH_S \over {d\tau}}\right\rangle_{rf} = \sum_b
\Gamma_{ab}^{rf}(\omega_{ab}), \label{eq30}\eeq
where we have used Eqs. (\ref{eq17}) and (\ref{eq19}).
Following essentially the same procedure,
an equation analogous to (\ref{eq31}) can be derived for the contribution
of self reaction to the energy shift
\beq \left( \delta E_a \right) _{sr}= {1 \over {2 \pi}}\sum_b
  \int_{-\infty}^{+\infty} d\omega' \,  {\Gamma_{ab}^{sr}(\omega') \over
  {\omega'}}   {{\cal P} \over
  {\omega' - \omega_{ab}}} ,  \label{eq32}\eeq
where
\beq \Gamma_{ab}^{sr}(\omega') :=
  2ig^2\sum_{i,j}\int_{0}^{\infty}d\tau' \, \chi_{ij}^R(x(\tau),x(\tau'))
  {d \over {d\tau}} \hbox{Re}\left(\langle a|S_i^f(0)
  |b \rangle \langle b|S_j^f(0)|a \rangle  e^{i\omega'(\tau - \tau')}\right)
  \label{eq32b}\eeq
and Eq. (\ref{eq30}) holds correspondingly (replace $rf$ by $sr$).
Note that $\Gamma_{ab}^{rf/sr}$ do not depend on $\tau$ because of the
stationarity of the physical situation.

We have studied above a very general stationary situation:
arbitrary linear system-reservoir coupling, arbitrary
stationary motion in a flat or curved spacetime, nontrivial boundary conditions
allowed, arbitrary state of the reservoir. We have shown that the
determination of energy shift and relaxation rate can be reduced
in a unified way
directly to the calculation of the coefficients $\Gamma_{ab}^{rf/sr}$.
They turn out to be the fundamental underlying quantities. The
calculation of the energy shift has thereby been simplified
as compared to (\ref{eq23}) and (\ref{eq24}). Equations
(\ref{eq31}), (\ref{eq30}) and (\ref{eq32}) show that the
$\Gamma_{ab}^{rf/sr}$ refer to particular transitions $|a \rangle \to
|b \rangle$. This becomes even more evident when the system has only a
few energy states as will be demonstrated in the next section.
In addition we see that the mutual dependence between energy shifts
and relaxation rates holds for the reservoir
fluctuation terms and for the self reaction terms separately. This
supports the view that the distinction of these two
mechanisms is physically
reasonable. Finally we mention that for concrete physical situations
it may be necessary to introduce a frequency cutoff in order to
regularize $(\delta E_a)_{rf/sr}$. As compared with (\ref{eq23})
and (\ref{eq24}), the new expressions (\ref{eq31}) and (\ref{eq32})
are directly prepared for this.

\section{Application: Two-level atom}

To demonstrate the usefulness of the proposed scheme we consider a
two-level system moving on a stationary trajectory and derive several
statements which are valid under very general conditions. This
illustrates at the same time the physical concepts.
We assume two stationary states $|+ \rangle$ and $|- \rangle$ with
energies $\pm {1\/ 2} \omega_0$. The atomic Hamiltonian can be written
with the help of the pseudospin operator $ S_3={1 \over 2}|+\rangle \langle+|
-{1 \over 2}|-\rangle \langle-| $ as
\beq H_S=\omega_0 S_3(\tau) \label{eq33}\eeq
The coupling to the reservoir is linear and connects only different levels
of the system:
\beq V= -g S_2(\tau) R(x(\tau)) \label{eq34}\eeq
with $ S_2={i \over 2}(S_+ - S_-) $ and $S_\pm = |\pm \rangle \langle \mp|$.
We rewrite $\Gamma_{ab}^{rf/sr}$ from Eqs. (\ref{eq31b}) and (\ref{eq32b}) as
\beq \Gamma_{ab}^{rf/sr}(\omega_{ab}) = - 2\omega_{ab} |\langle
  a|S_2^f(0)|b\rangle |^2 \gamma^{rf/sr}(\omega_{ab}) \label{eq36b}\eeq
where we have simply $|\langle a|S_2^f(0)|b\rangle |^2 = {1\/4}$ for
$a \neq b$. Here,
\beq \gamma^{rf}(\omega_{ab}) = g^2
  \int_{0}^\infty d \tau' C^R(x(\tau),x(\tau')) \cos \omega_{ab}
  (\tau-\tau'), \label{eq36a}\eeq
\beq \gamma^{sr}(\omega_{ab}) = i g^2
  \int_{0}^\infty d \tau' \chi^R(x(\tau),x(\tau')) \sin \omega_{ab}
  (\tau-\tau'), \label{eq37a}\eeq
and $\omega_{ab}=\pm \omega_0$ for $a>b$ ($a<b$).

Using the general expression (\ref{eq30}) for the relaxation rate together
with (\ref{eq36b}), we can write
\beq \left\langle {dH_S \over {d\tau}}\right\rangle_{rf} =
  - 2 \sum_b \omega_{ab} |\langle
  a|S_2^f(0)|b\rangle |^2 \gamma^{rf}(\omega_{ab}) \label{eq35}\eeq
Since $\gamma^{rf}$ is symmetric in $\omega_{ab}$, it follows that
\beq \left\langle {dH_S \over {d\tau}})\right\rangle_{rf} = -2\omega_0
  \gamma^{rf}(\omega_0) \left(
  \sum_{b<a} |\langle a|S_2^f(0)|b\rangle|^2
  + \sum_{b>a} |\langle a|S_2^f(0)|b\rangle |^2 \right)\label{eq36}\eeq
Analogously, we find for the contribution of self reaction to the
relaxation rate
\beq \left\langle {dH_S \over {d\tau}}\right\rangle_{sr} = -2\omega_0
  \gamma^{sr}(\omega_0) \left(
  \sum_{b<a} |\langle a|S_2^f(0)|b\rangle |^2
  -\sum_{b>a} |\langle a|S_2^f(0)|b\rangle |^2 \right). \label{eq37}\eeq

{}From these general relations, it is possible to find expressions for
the atom's evolution into equilibrium as well as for the Einstein
coefficients $A_\uparrow$ and $A_\downarrow$ corresponding to upwards
and downwards transitions.  We consider the total transition rate
$\left\langle {dH_S \over {d\tau}} \right\rangle_{tot} =\left\langle {dH_S
\over {d\tau}} \right\rangle_{rf} + \left\langle {dH_S \over {d\tau}}
\right\rangle_{rf}$. Using (\ref{eq36}) and (\ref{eq37}), it can be
simplified by noting
\beq  \sum_{b>a} |\langle a|S_2^f(0)|b\rangle |^2
   \pm \sum_{b<a} |\langle a|S_2^f(0)|b\rangle |^2
   =\cases{{1 \over 4}&\cr -{ 1\/ 2\omega_0}  \langle a|H_S|a\rangle. &\cr}
   \label{eq38}\eeq
Here we have replaced $H^f_S$ by the total system Hamiltonian, which
is justified in order $g^2$.
We thus obtain a differential equation for the evolution of the mean atomic
excitation energy
\beq \left\langle {dH_S \over {d\tau}}(\tau)\right\rangle_{tot} = -{1 \over 2}
  \omega_0 \,\gamma^{sr}(\omega_0) - \gamma^{rf}(\omega_0) \,
  \langle H_S(\tau) \rangle \label{eq39}\eeq
which has the solution
\beq \langle H_S(\tau) \rangle= -{1\/ 2}\omega_0 +{1\/ 2}\omega_0
  {\gamma^{rf}-\gamma^{sr} \/ \gamma^{rf}} +\left(\langle H_S(0) \rangle
  +{1\/ 2}\omega_0 {\gamma^{sr}\/ \gamma^{rf}}\right) e^{-\gamma^{rf}\tau}
  \label{eq40}\eeq
A number of interesting points can be inferred from (\ref{eq40}):
First the relaxation into equilibrium is always exponential, without
oscillations, no matter what the the trajectory of the atom or the state
of the
reservoir is. Its rate is determined by the contribution of reservoir
fluctuations $\gamma^{rf}$ alone. Furthermore, the first two terms
of (\ref{eq40}) show that the equilibrium excitation above the ground
state $-{1\/ 2}\omega_0$ is governed by the relative magnitude of
$\gamma^{rf}$ and $\gamma^{sr}$.

The Einstein coefficients $A_\uparrow$ and $A_\downarrow$ can be
identified by comparison of (\ref{eq40}) with an appropriate rate
equation (Eq. (65) of \cite{Audretsch94a}). One finds
\beq A_\uparrow = {1\/2}\gamma^{rf} - {1\/2}\gamma^{sr}, \qquad
  A_\downarrow = {1\/2}\gamma^{rf} + {1\/2}\gamma^{sr}.
  \label{eq47a}\eeq
The important consequence is that $\gamma^{rf}$ and $\gamma^{sr}$
(and therefore $\Gamma^{rf}_{ab}$ and $\Gamma^{sr}_{ab}$) can separately
be determined as functions of the Einstein coefficients. In principle
they are therefore measurable quantities. Beyond that one can read
off from (\ref{eq47a}) the physical origin of the Unruh effect
\cite{Unruh76}. Taking as reservoir the vacuum, the spontaneous excitation
of an accelerated two-level atom
results from an imbalance between the contributions of vacuum fluctuations and
radiation reaction as has been discussed in detail in \cite{Audretsch94a}.

The radiative energy shift, which in this case is called the Lamb shift
of the two-level atom can
now be obtained by application of the formulas (\ref{eq31}) and (\ref{eq32}).
First we notice that generally the contribution of self reaction
does not contribute to the relative shift of the two levels. From
(\ref{eq32}) it follows after a short calculation that
\beq \Delta_{sr} = \delta E_+^{sr} - \delta E_-^{sr}= 0. \eeq
On the other hand, the contribution of reservoir fluctuations is
\bea \Delta_{tot} &=& \Delta_{rf}=\delta E_+^{rf} - \delta E_-^{rf} \nn\\
&=& {1 \over {2\pi}} \int_0^{\infty} d\omega' \, \gamma^{rf}
  (\omega') \left( {{\cal P} \over {\omega' + \omega_{0}}} -
  {{\cal P} \over {\omega' - \omega_{0}}} \right), \label{eq47}\eea
where the symmetry properties of $\gamma^{rf} (\omega')$ have been used.
It is determined by the fluctuations of the reservoir in accordance with the
heuristic picture of Welton \cite{Welton48}. In the derivation of
Eq. (\ref{eq47}), no other properties
of the reservoir and no further assumptions about boundaries or the
atomic trajectory were used. Their influences are contained in the form
of $\gamma^{rf}$. Note that the energy shift can be determined once
the functional form of the Einstein coefficients (\ref{eq47a}) is
known. This is so much the more remarkable since, in contrast to the
usual expression for the Lamb shift, they can be calculated easily
for a given system, for example from Fermi's golden rule.


\end{document}